\documentclass[11pt]{article}

\usepackage[utf8]{inputenc}
\usepackage{amsmath}
\usepackage{amsfonts}
\usepackage{amssymb}

\usepackage{caption}
\usepackage{graphicx}
\usepackage{epstopdf}

\newcommand{\ie}{\emph{i.e.}\ }
\newcommand{\cf}{cf.\ }

\newcommand{\B}{\mathcal{B}}

\title{Self-ordering and collective dynamics of transversely illuminated point-scatterers in a 1D trap}
\author{Daniela~Holzmann, Matthias~Sonnleitner and Helmut~Ritsch\\
Institute for Theoretical Physics\\ University of Innsbruck \\  A-6020 Innsbruck, Austria}

\date{\today}

\begin{document}

\maketitle

\begin{abstract}
 We study point-like polarizable particles confined in a 1D very elongated trap within the evanescent field of an optical nano-fiber or nano-structure. When illuminated transversely by coherent light, collective light scattering into propagating fiber modes induces long range interactions and eventually crystallisation of the particles into regular order. We develop a simple and intuitive scattering-matrix based approach to study these long-range interactions by collective scattering and the resulting light-induced self-ordering. For few particles we derive explicit conditions for self-consistent stable ordering. In the purely dispersive limit with negligible back-scattering, we recover the prediction of an equidistant lattice as previously found for effective dipole-dipole interaction models. We generalize our model to experimentally more realistic configurations including backscattering, absorption and a directional scattering asymmetry. For larger particle ensembles the resulting self-consistent particle-field equations can be numerically solved to study the formation of long-range order and stability limits.
\end{abstract}
\section{Introduction}

Optical control and manipulation of atoms and nano-par\-ticles in free space has seen tremendous progress in the past decade and allows for cooling and trapping in designable optical traps of almost any shape. In particular, precise periodic optical potentials can be generated, which allow one to study important solid state lattice Hamiltonians. In contrast to conventional solids, however, the spatial order and lattice geometry is perfect and fixed by the external lasers. Any back-action of the particles on the light and corresponding long-range interactions are generally neglected, but would lead to very interesting new dynamical effects~\cite{deutsch1995photonic,asboth2008optomechanical}.

In an important step Rauschenbeutel and coworkers, managed to trap atoms in an array of optical dipole traps generated by two-colour evanescent light fields alongside a tapered optical fiber~\cite{vetsch2010optical}, where the back-action of even a single atom on the propagating fiber field is surprisingly strong~\cite{domokos2002quantum}. This setup was improved with higher control and coupling by other groups recently~\cite{goban2012demonstration,lee2013integrated}. With the atoms firmly trapped within the evanescent modes, field mediated atom-atom interaction and collective coupling to the light modes play a decisive role in this setup~\cite{zoubi2010hybrid}. Interestingly trapping, ordering via optical binding and collective forces on nano-particles trapped by optical nano-fibers have also been recently observed using only a single laser~\cite{frawley2014selective}.  

For cold particles in optical resonators transversely illuminated by lasers it is well established by now, that a phase transition from a homogeneous to  regular order appears at a sufficiently strong pump intensity and suitable detunings between laser and cavity field~\cite{asboth2005self,chan2003observation}.  In this work we will explore self-ordering effects of such a transverse laser beams on atoms trapped near a nano-fiber.

As a quite surprising generalisation Chang and coworkers predicted in a recent theoretical model that dipole-dipole interaction via the fiber mode can induce a stable regular order and nontrivial long-range correlations in such a nano-fiber geometry~\cite{chang2012cavity}. In a closely related approach, we have theoretically studied a generalised model of this setup based on the effective coupled Vlasov equation for the atomic phase space distribution and the field distribution in the fiber~\cite{griesser2013light}. This model predicted threshold conditions for the appearance and stability of light scattering induced density modulations of the particle distribution at finite temperature. \\Depending on pump strength and particle number several nontrivial steady states appear, in which multiple light scattering by the particles confines the light modes. At the same time the density modulations of the particle distribution is sustained by the light in the fiber.

In this work we study the microscopic origin and coupling dynamics of this dynamic self-ordering using a scattering formalism to calculate the field amplitudes and for\-ces acting on each individual particle. While this approach is less suitable to study thermodynamic properties of large ensembles, it gives a very intuitive picture of the underlying physics and can be easily applied to a variety of different situations from single atoms, molecules to larger nano-particles trapped in vacuum or in a viscous transparent medium. Here we start from explicitly analytically solvable few particle cases, where the underlying microscopic physical processes can be clearly identified. For larger particle-numbers numerical simulations of the self-ordering process and the appearance of order can be studied in detail. Generalisations to directionally biased scattering into the fiber and transverse multimode fibers are easily possible within this approach.   

\begin{figure}
 \centering
 \includegraphics[width=1\columnwidth]{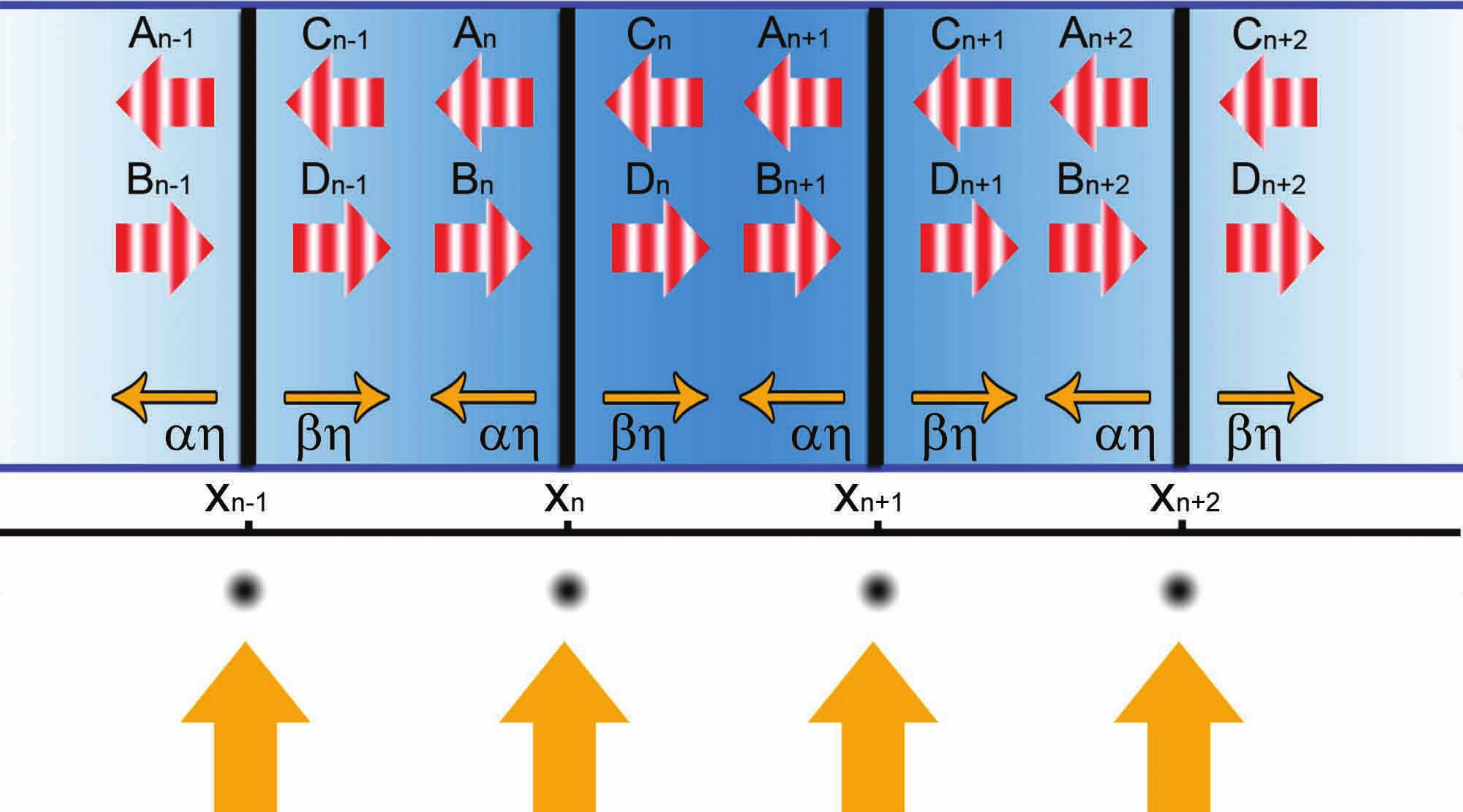}
 \caption{A 1D array of point particles scattering light in and out of an optical nano-structure can be modelled as a collection of beam splitters interacting with a plane wave.}
\label{fig_system}
\end{figure}%

In the following, after introducing our model in terms of a simple self-consistent scattering-matrix approach, we study the case of a single particle subject to external pump and the fields induced by neighbouring scatterers and derive the conditions for a stable equilibrium point. Next we consider two and three particle configurations, where fairly simple analytical results still can be obtained and the essential physics of trapping and self-ordering can be seen in detail. Numerical simulations of particle motion for larger ensembles then allow us to find stability conditions in the last part of this work.

\section{Scattering model of particle-field dynamics}
Let us consider $N$ polarizable particles confined in a 1D potential parallel to a tapered fiber as depicted in Fig.~\ref{fig_system}. Via coupling to the evanescent field of the fiber mode, the particles can scatter photons into and out of the two propagating light field modes~\cite{chang2012cavity,griesser2013light,mitsch2014discerning}. The fiber modes can be described by two counter-propagating fields of frequency $\omega=k c$. For a particle at position $x_j$ the fields left and right of this particle shall be written as
\begin{equation}\begin{split}
	E_l(x)&=A_j \exp(-i k (x-x_j)) + B_j \exp(i k (x-x_j)), \\
	E_r(x)&=C_j \exp(-i k (x-x_j)) + D_j \exp(i k (x-x_j))
\end{split}\end{equation}
for $j=1,\dots,N$. The perturbation induced by the particles induces scattering between the incoming fields $B_j$ and $C_j$ and outgoing fields $A_j=r B_j+t C_j$ and $D_j = t B_j+r C_j$, the particle therefore acts as an effective beam splitter with prescribed reflection and transmission coefficients $r$ and $t$.

In addition each particle scatters light from a transverse pump laser into the fiber modes. The exact treatment of how light is scattered into a nano-fiber by transversely illuminated atoms is very complex and depends on various parameters such as the polarization of the modes or particle properties~\cite{mitsch2014discerning}. Here we make the simplifying assumption that the scattering process can be expressed by a single effective amplitude~$\eta$ which is the same for all~$N$ particles. From the perspective of the fiber modes, these particles then not only act as beam splitters, but also as sources. If we allow for asymmetric scattering we may therefore write for the field amplitudes 
\begin{equation}
	A_j=r B_j+t C_j + \alpha \eta, \qquad
	D_j = t B_j+r C_j+\beta \eta,
\end{equation}
for all $j=1,\dots,N$ and with $\alpha^2+\beta^2=1$, $\alpha, \beta \in \mathbb{R}_{\geq 0}$. In analogy to the scattering model for optical lattices~\cite{asboth2008optomechanical,sonnleitner2012optomechanical,ostermann2014scattering} we can write a generalised scattering matrix for a single particle with coupling constant $\zeta=k \tilde{\alpha}/(2\epsilon_0)$, with the effective polarizability $\tilde{\alpha}$ describing the interaction between the particle and the fiber mode. The transmission and reflection coefficients are then given by $t=1/(1-i\zeta)$ and $r=i\zeta/(1-i\zeta)$, respectively. 

The coupling between the amplitudes left and right of a particle can then be expressed using the following scattering matrix
\begin{equation}
\begin{split}
	\begin{pmatrix} A_j\\B_j\\ \eta \end{pmatrix} 
		&= \frac{1}{t}\begin{pmatrix} t^2-r^2 & r & \alpha t-\beta r \\ -r & 1 & -\beta \\ 0 & 0 & t  \end{pmatrix}
				\begin{pmatrix} C_j\\D_j\\ \eta \end{pmatrix} \\
		&= \begin{pmatrix} 1+i\zeta & i\zeta & \alpha-i\beta\zeta \\ -i\zeta & 1-i\zeta & \beta(i\zeta-1) \\ 0 & 0 & 1  \end{pmatrix} 
				\begin{pmatrix} C_j\\D_j\\ \eta \end{pmatrix} 
%				= M\begin{pmatrix} C_j\\D_j\\ \eta \end{pmatrix}
\end{split}
\end{equation}
where $B_j,C_j,\eta$ are the incoming fields allowing to determine the outgoing fields $A_j,D_j$, \cf Fig.~\ref{fig_system}. These then in turn will be used as inputs to neighbouring scatterers as $C_{j-1}=A_j \exp(-i k (x_j-x_{j-1}))$ and $D_{j-1}=B_j \exp(i k (x_j-x_{j-1}))$.  Note that we recover the well established transfer-matrix model~\cite{asboth2008optomechanical,sonnleitner2012optomechanical,ostermann2014scattering,xuereb2009scattering} for an array of particles interacting with plane waves, if we turn off the transverse pump by setting $\eta=0$.

In this work we are interested in the mechanical effects of the transverse laser beam on the motion of a one-dimensional chain of beam splitters along the fiber.  The force arises from the scattering of the transverse field into the fiber, which will interfere with existing fields and in particular with the light scattered by other particles. This gives versatile dynamics along the $x$-direction including a new kind of optical binding and self-ordering.

In addition the pump laser can act in the transverse direction as well and modify the distance between the particles and the fiber. Here we assume that this has only a minor effect on the trap. 

Using simple arguments based on the Maxwell stress tensor, the time averaged force along the $x$-axis on the $j$-th particle is given by~\cite{asboth2008optomechanical}
\begin{equation}\label{eq_force_BSj_general}
	F_j=\frac{\epsilon_0}{2}\left(\vert A_j\vert^2+\vert B_j\vert^2-\vert C_j\vert^2-\vert D_j\vert^2\right).
\end{equation}
This simple expression for the local force tends to shroud the inherent complexity of this system. We will start to examine this rich behaviour by a detailed study of the forces on a single particle interacting with longitudinal and transverse light-fields.

\section{Scattering force on a single particle}

As presented above a single particle acts as a scatterer, which couples the left and right propagating modes and in addition coherently scatters pump light into these modes. The local field amplitudes $B_1$ and $C_1$  at $x=x_1$ are determined by intensities and phases of the longitudinal beams via $B_1=\sqrt{2 I_l/(c \epsilon_0)} \exp(i k x)$, $C_1=\sqrt{2 I_r/(c \epsilon _0)} \exp(-i k x)$ and the scattering amplitude originating from the transverse pump is given by $\eta = \sqrt{2 I_\eta/(c \epsilon_0)} \exp(-i \phi)$, with a factor $\phi$ describing the phase difference between the longitudinal and the effective transverse pump fields. Using Eq.~\eqref{eq_force_BSj_general}, the most general expression for the force in terms of the light field intensities then reads
\begin{equation}\label{eq_force_singleBS_general}
	\begin{split}
		F&=\frac{(I_l-I_r)\left(\left|\zeta\right|^2+\zeta_i\right)-2\sqrt{I_l I_r}\zeta_r \sin(2 k x)}{\frac{c}{2}\vert 1-i\zeta\vert^2}\\
		&+\frac{I_{\eta}}{c}(\alpha^2-\beta^2)\\
		&+\frac{2\sqrt{I_\eta I_l}}{c} \Re\left(\frac{i \alpha\zeta-\beta}{1-i\zeta}e^{i (k x+\phi)}\right)\\
		&+\frac{2\sqrt{I_\eta I_r}}{c} \Re\left(\frac{\alpha-i\beta\zeta}{1-i\zeta}e^{-i (k x-\phi)}\right).
	\end{split}
\end{equation}
The first line represents the optical lattice generated by the counter propagating fields in the fiber~\cite{asboth2008optomechanical} and the second line represents the radiation pressure force induced by asymmetric scattering of the transverse pump light into the fiber if $\alpha \neq \beta$. The last two lines emerge from interference between scattered and longitudinal fields with $\zeta_r=\Re(\zeta)$ and $\zeta_i=\Im(\zeta)$ and are of central importance for long range interactions and self-ordering. 

Equation~\eqref{eq_force_singleBS_general} already contains all parameters that determine the dynamics in this surprisingly complex system: the intensities of the longitudinal and transverse pump beams,~$I_l, I_r$ and $I_\eta$; the relative phase between those beams $\phi$, the effective coupling between the particles and the modes travelling the nanofiber $\zeta$ and the coefficients giving the scattering asymmetry $\alpha$ and $\beta$~\cite{mitsch2014directional}.

For symmetric scattering $\alpha = \beta = 1/\sqrt{2}$ we have
\begin{equation}\label{F1}
	\begin{split}
		F&=\frac{(I_l-I_r)\left(\left|\zeta\right|^2+\zeta_i\right)-2\sqrt{I_l I_r}\zeta_r \sin(2 k x)}{\frac{c}{2}\vert 1-i\zeta\vert^2}\\
		&+\frac{\sqrt{2 I_\eta}}{c} \left(\sqrt{I_r}\cos(kx-\phi)-\sqrt{I_l}\cos(kx+\phi)\right),
	\end{split}
\end{equation}
which for equally strong left and right propagating fields $I_r=I_l=I$ reduces to:
\begin{equation}
\begin{split}
F&=\frac{2 \sin(k x)}{c}\left(\frac{-4 I\zeta_r \cos(k x)}{\vert 1-i\zeta\vert^2}+\sqrt{2 I_\eta I}\sin(\phi)\right).
\end{split}
\end{equation}

\begin{figure}
\centering
	\includegraphics[width=\columnwidth]{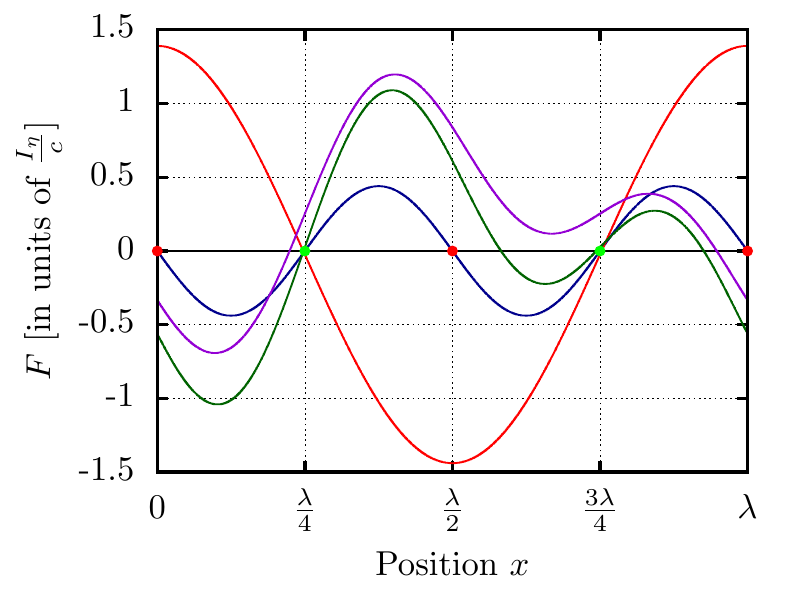} 
	\caption{Force on a single beam splitter as function of the position of the beam splitter $x$ for $\alpha=\beta$ and $\phi=0$. The blue line corresponds to $I_l=I_r=I_\eta$ and $\zeta=1/9$, the red line to $I_l=0$, $I_r=I_\eta$ and $\zeta=1/9$, the green line to $I_l=2 I_\eta$, $I_r=I_\eta$ and $\zeta=1/9$ and the violet line to $I_l=2 I_\eta$, $I_r=I_\eta$ and $\zeta=(1+i)/9$. Stable trapping points with $F=0$ are marked by red dots, unstable by green dots.}
	\label{fig:f1a}
\end{figure}%

This form clearly separates the lattice and transverse pump contribution. Obviously without transverse pump, $I_\eta=0$, we get a simple standing wave lattice with period $\lambda/2$. Adding a transverse pump $I_\eta$ we observe a transition from the $\lambda/2$-periodicity to $\lambda$-periodicity as for cavity induced self-ordering~\cite{asboth2005self}. A finite light absorption rate is parametrised by the imaginary part of $\zeta$ and adds a radiation pressure force shifting the lattice constant like in a conventional optical lattice (Fig.~\ref{fig:f1b})~\cite{asboth2008optomechanical}. Note that the force on a single particle vanishes without propagating longitudinal fields $I_l=I_r=0$. 

The typical spatial dependence of the force presented in Fig.~\ref{fig:f1a} for different parameter sets exhibits one or several stable equilibrium positions, which correspond to zeros of the force with negative gradient. Here they can be explicitly determined for $I_l=I_r=I$ as: 
\begin{equation}\label{kxIrIl}
	kx=\begin{cases}
		2n\pi, ~~~~~~~~~~~~~~~~~~~~~~~~\text{ if  }\sin(\phi)<\frac{2\sqrt{2 I}\zeta_r}{\sqrt{I_\eta}\vert 1-i\zeta\vert^2},\\
		(2n+1)\pi,~~~~~~~~~~~~~~~~~\text{ if  }\sin(\phi)>\frac{-2\sqrt{2 I}\zeta_r}{\sqrt{I_\eta} \vert 1-i\zeta\vert^2},\\
		\pm\arccos\left(\sqrt{\frac{I_\eta}{I}}\frac{\vert 1-i\zeta\vert^2}{2\zeta_r}\sin(\phi)\right)+2n\pi, \text{   if  }\zeta_r<0,\\
	\end{cases}
\end{equation}

Note that for symetric pump $I_r=I_l$ and $\phi=0$ as for the blue line in Fig.~\ref{fig:f1a}, the second line of Eq.~\eqref{F1} for the force vanishes as for $I_\eta=0$ and we get a simple standing wave lattice with $\lambda/2$-periodicity. Choosing $I_l\neq I_r$  one eventually looses this periodicity and only one stable position per wavelength survives. Adding an imaginary part of $\zeta$  shifts the force zeros towards the weaker source.

\begin{figure}
	\includegraphics[width=\columnwidth]{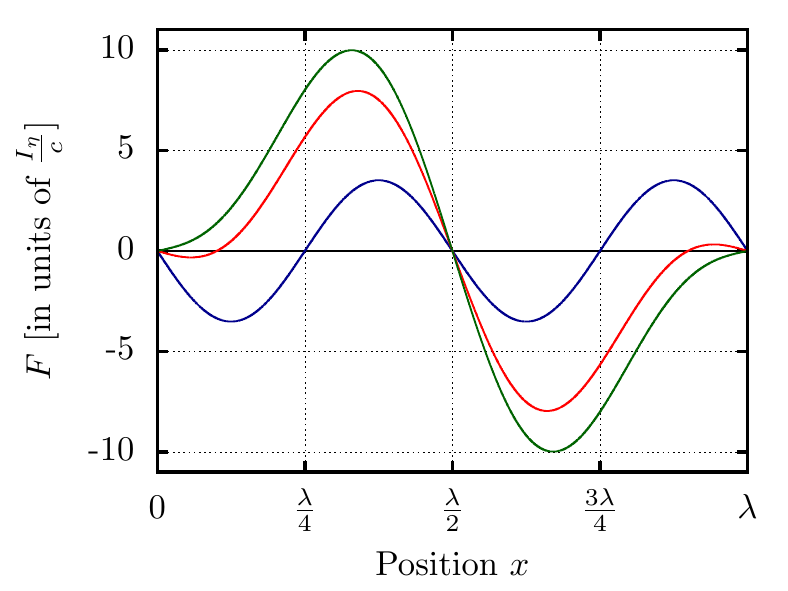}
	\caption{Force on a single beam splitter for $I_l=I_r=8 I_{\eta}$, $\zeta=1/9$ and $\alpha=\beta$. The blue line corresponds to $\phi=0$, the red line to $\phi=\pi/4$ and the green line to $\phi=\pi/2$. We see that the particle trap positions at integer multiples of $\lambda$ become unstable for $\phi>\pi/4$, cf. equation~\eqref{kxIrIl}.}
	\label{F1eta}
\end{figure}

\begin{figure}
	\includegraphics[width=\columnwidth]{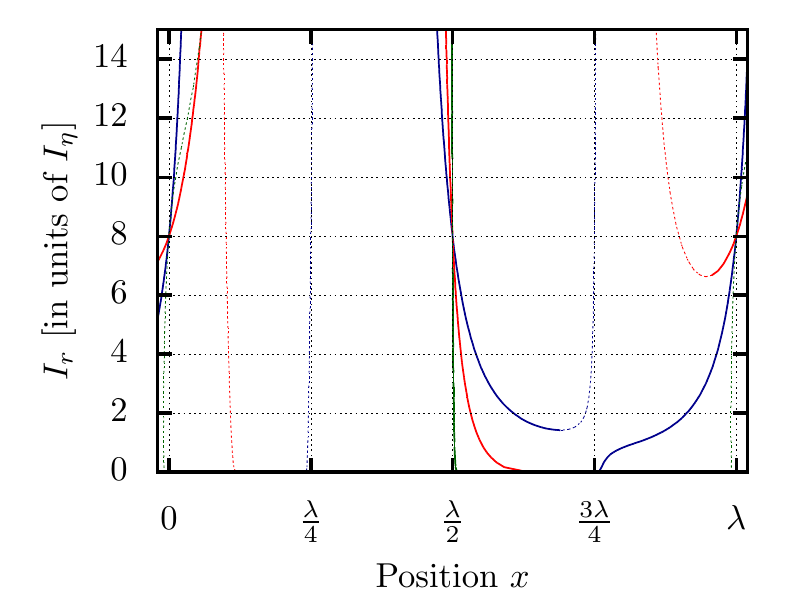} 

	\caption{Dependence of the zero force position $x$ for varying pump right field power $I_r$ for $\alpha=\beta$, $\zeta=1/9$ and $I_l=8 I_\eta$. The blue line corresponds to $\phi=0$, the red line to $\phi=\pi/4$ and the green line to $\phi=\pi/2$. The solid lines show the stable zero-points, while the dashed lines show the unstable ones.  Comparing this figure with Fig.~\ref{F1eta} we see that the figures correspond to each other at $I_r=8 I_\eta$. It is interesting to see that for some $\phi$'s new zero-points appear after exceeding a special threshold.}
	\label{F1etairc}
\end{figure}

The solutions given by the last line of Eq.~\eqref{kxIrIl} only exist for $\vert\sin(\phi)\vert$ $ \leq 2\sqrt{2 I}{\zeta_r}/(\sqrt{I_\eta} \vert 1-i\zeta\vert^2)$. Typical examples for the dependence of the force on the particle on the phase of the pump are shown in Fig.~\ref{F1eta}. Choosing $\zeta_r>0$ we see that the first stationary position at $x=0$ (or $x=\lambda$) is only stable as long as additional unstable zeros at 
\begin{equation}
kx=\pm\arccos\left(\sqrt{I_\eta/I}\frac{\vert 1-i\zeta\vert^2}{2 \zeta_r}\sin(\phi)\right)+2n\pi
\end{equation}
(Eq.~\eqref{kxIrIl}, condition 3) exist too. In general  one can get even more zero-force points per wavelength for larger ranges of  $I_\eta$, $I$, $\phi$ for $\zeta$ as depicted in the example of Fig.~\ref{F1etairc}. 

%\begin{figure}
%\include{F1etac}
% \caption{Zeropoints of the force on one beam splitter for $I_l=I_r=I$, $\zeta=\frac{1}{9}$ and $\theta=\frac{\pi}{4}$. The blue line corresponds to $\phi=0$, the red line to $\phi=\frac{\pi}{4}$, the green line to $\phi=\frac{\pi}{3}$ and the violet line to $\phi=\frac{\pi}{2}$.}
% \label{F1etac}
%\end{figure}

\subsection{Stability of the outermost particle in a chain} 
Let us discuss here another special case, which will later be important for the multiparticle case. In a generic setup, where only a transverse pump laser is present, all propagating photons originate from scattering by the particles. Hence the outermost particles will only be exposed to incoming light from one side. The impinging amplitude then corresponds to the total light scattered by all other particles into its direction. As any spatial configuration can only be stable, when the outermost particle is also stationary, it is useful to study the stability conditions for such single-side illumination first.

Setting $I_r=0$ in Eq.~\eqref{F1} the force on the rightmost particle reads
\begin{equation}
	F=\frac{1}{c}\left(\frac{2 I_l\left(\left|\zeta\right|^2+\zeta_i\right)}{\vert 1-i\zeta\vert^2}-\sqrt{2 I_\eta I_l}\cos(kx+\phi)\right)
\end{equation}
with $\zeta_i=\Im(\zeta)$. We see that the interference term in this equation can compensate for the radiation pressure generated from the incoming light from the left. Indeed the total force vanishes for
\begin{equation}
k x=-\arccos\left(\frac{\sqrt{2}(\zeta_i+\vert\zeta\vert^2)}{\vert 1-i\zeta\vert^2}\sqrt{\frac{I_l}{I_\eta}}\right)-\phi+2n\pi, n\in\mathbb{N}
\end{equation}

\begin{figure}
	\includegraphics[width=\columnwidth]{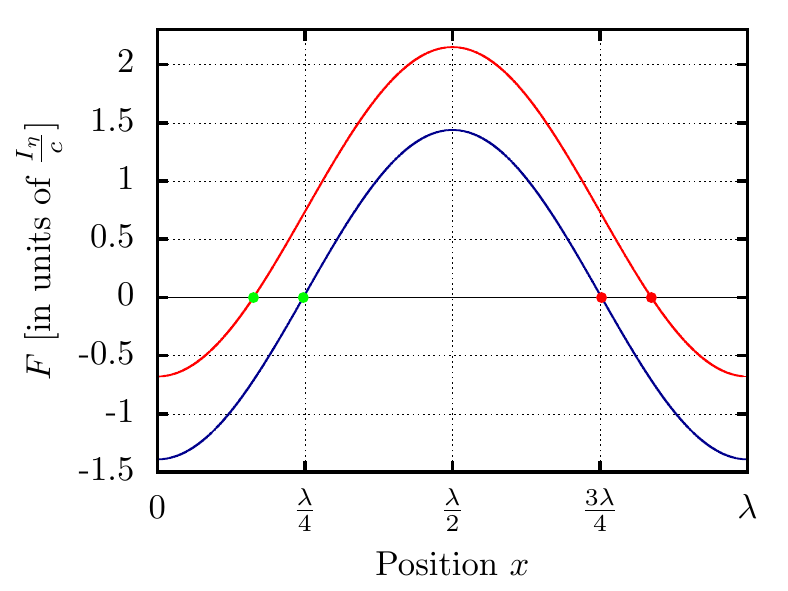}
	\caption{Force on a single particle illuminated just from the left as $I_l=I_\eta$ and $I_r=0$; $\phi=0$, $\alpha=\beta$. The blue line corresponds to $\zeta=1/9$ and the red line to $\zeta=1/9+i/2$. The red points are stable and the green ones not. Since $I_r<I_r$ and $\zeta_i\neq 0$ we see a radiation pressure force pushing the particle towards the right.}
	\label{fig:f1b}
\end{figure}

Examining this expression we can see that we can only find a zero force position, if $I_l$, $I_\eta$ and $\zeta$ fulfil:
\begin{equation}
	\frac{I_\eta}{I_l}\geq \left(\frac{\sqrt{2}(\zeta_i+\vert\zeta\vert^2)}{\vert 1-i\zeta\vert^2}\right)^2,
\end{equation}
i.e. $I_\eta/I_l$ has to exceed a certain threshold. As $I_l$ is proportional to $I_\eta$ and depends on the particle number, this condition eventually limits the maximal particle number which can form a stable chain. Unfortunately the explicit expressions for $I_l$ are rather complex, so that a simple stability criterion is difficult to obtain. Nevertheless, as depicted in Fig.~\ref{fig:f1b}, we find stable zero force positions for the last particles even when absorption strongly dominates scattering (red dot on red curve).  

\section{Collective scattering and forces for several particles}

For $N>1$  particles the motion is conveniently split into center of mass (CMS) and relative motion. Hence besides completely stationary solutions, where all particles are at rest, we can find cases of an equal nonzero force on all particles, so that they move together at a fixed distance. The condition of a stationary center of mass can be simply expressed in terms of the outermost field amplitudes to give
\begin{equation}
F_{tot}=\frac{\epsilon_0}{2}(\vert A_1\vert^2+\vert B_1\vert^2-\vert C_N \vert^2-\vert D_N\vert^2)=0,
\end{equation}
for any particle number $N$.

\subsection{Two particles}
Let us first look at two particles at a given distance $d$ without injected fields $I_l=I_r=0$ and symmetric scattering $\alpha=\beta=1/\sqrt{2}$. Here the light scattered into the fiber by one particle only interferes with the scattered light of the second particle. Evaluating the general expression Eq.~\eqref{eq_force_BSj_general} thus leads to a distance-dependent force, which for constructive interference induces a strongly attractive force and gives a repulsive term for destructive interference. Very close particles thus attract each other, while they repel at about half wavelength distance. Due to the negligible damping of the light propagation in the fiber this behaviour is periodically repeated over several wavelengths. To enable the comparison of the cases with more particles we normalize $I_{\eta_{tot}}=N I_\eta$.

\begin{figure}
\includegraphics[width=\columnwidth]{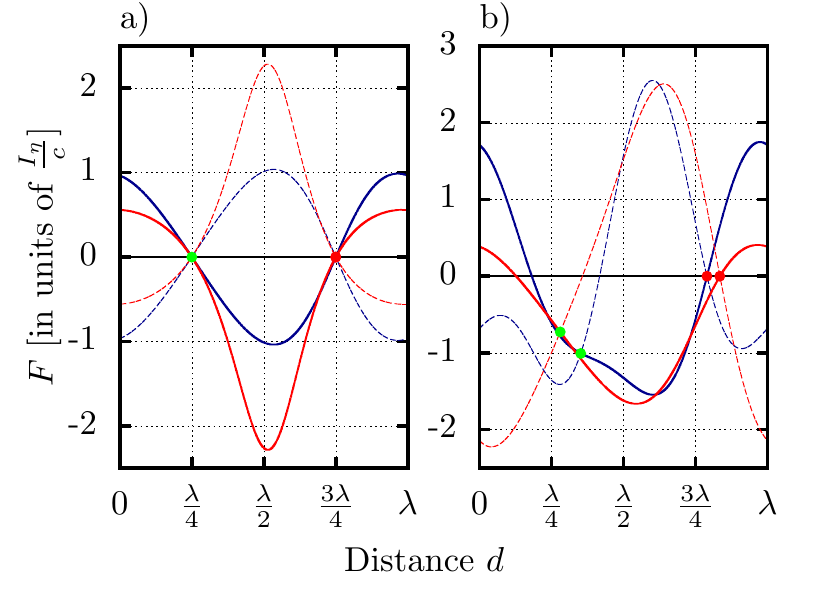}
\caption{Forces on two beam splitters as function of distance $d$ for $\alpha=\beta$ and $\phi=0$. The solid line shows the force on the first beam splitter and the dashed line the force on the second one. The red points are stable equilibria and the green ones not.
Left figure: The blue line corresponds to $I_l=I_r=0$ and $\zeta=1/9$, the red line to $I_l=0$, $I_r=0$ and $\zeta=1/9+i/2$. Because of symmetry $F_1=-F_2$.
Right figure: The blue line corresponds to $I_l=3 I_\eta$, $I_r= I_\eta$ and $\zeta=1/9$, the red line to $I_l=0$, $I_r= I_\eta$ and $\zeta=1/9$. Here the parameters are chosen to get zero center of mass force at the stationary distance. }
\label{force2}
\end{figure}

Explicitly for the total outgoing light intensity at both sides we get the superposition of light scattered by the two point particles exhibiting a periodic interference behavior  
\begin{equation} \label{eq}
 I_{ol}=I_{or}=2 I_\eta\left|\frac{(1-i\zeta) \cos (\frac{k  d}{2})}{(1-2i \zeta) \cos (\frac{k  d}{2})-i\sin (\frac{k  d}{2})}\right|^2
\end{equation}
with $I_{ol}=c \epsilon_0\vert A_1\vert^2/2$ and $I_{or}=c \epsilon_0\vert D_2\vert^2/2$. Symmetry here immediately implies equal forces of opposite sign on the two scatterers (\ie $F_1=-F_2$) with
\begin{equation}
	F_1=\frac{I_\eta \vert 1-i\zeta\vert^2 \cos (k d)}{c \left( 4 \left(\vert\zeta \vert^2+\zeta_i\right) \cos^2(\frac{k  d}{2})+ 2 \zeta_r \sin (k  d)+1\right)}.
\end{equation}
Stable distances with zero force on both particles are thus found at
\begin{equation}\label{d2}
	d=\left(\frac{3}{4}+n\right)\lambda
\end{equation}
for $n\in \mathbb{N}$. Fig.~\ref{force2} shows some examples for the force on two particles for different parameters. Configurations are stable, if the derivative of the force with respect to distance on the first beam splitter is positive and negative on the right beam splitter. Note that without injected lattice we get a stable distance of $d = 3 \lambda/4$ almost independent of the imaginary part of $\zeta$. This is exactly the distance where the scattered fields from the two particles are $90$~degrees out of phase and thus do not interfere. This distance is very different to cavity induced self-ordering, where maximal collective scattering at exactly wavelength distance leads to the most stable configurations~\cite{asboth2005self}. Adding extra fields forming an optical lattice via field injection through the fiber, the stationary distance of the particles changes and now depends more strongly on absorption as depicted in Fig.~\ref{force2}b).

\begin{figure}
  \includegraphics[width=0.8\columnwidth]{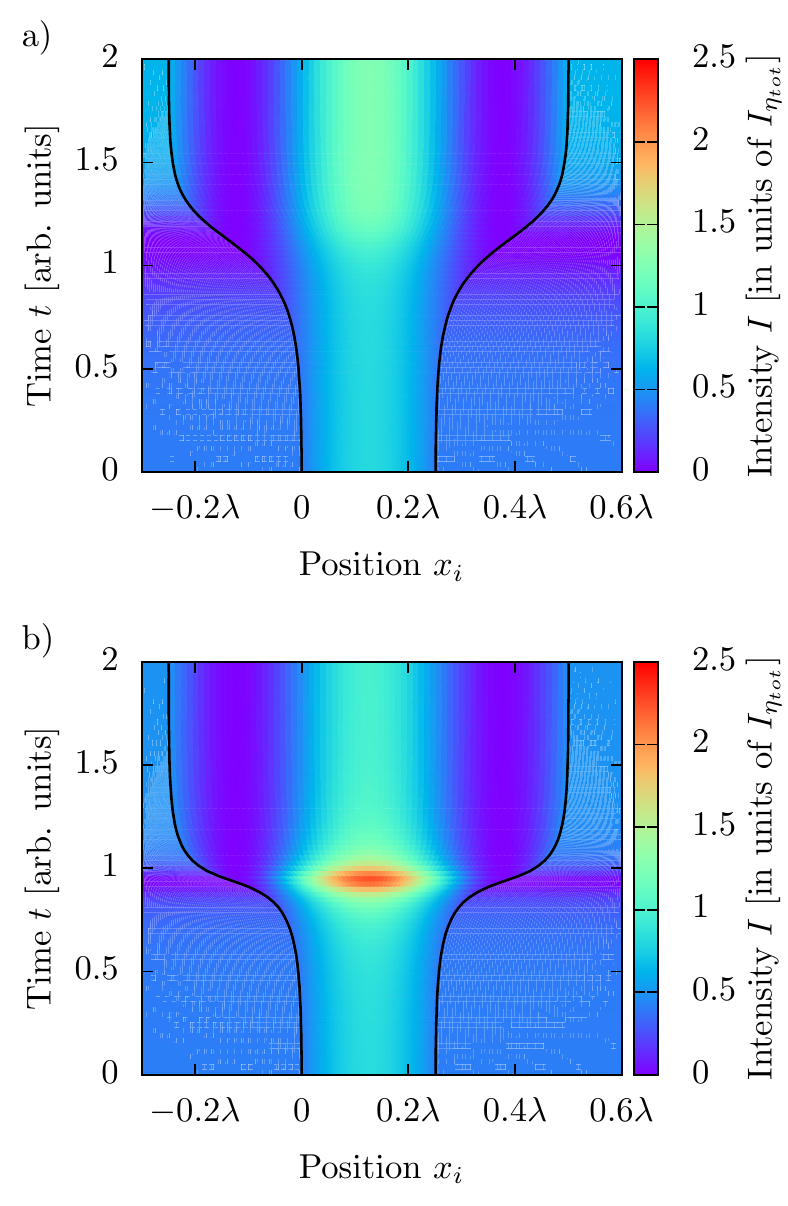}
  \caption{Trajectories and intensities of two beam splitters for $I_l=I_r=0$, $\alpha=\beta$ and $\phi=0$ with initial condition $d=\lambda/4$. Figure a) shows the trajectories for $\zeta=1/9$ and figure b) for $\zeta=1/9+i/2$. As calculated in Eq.~\eqref{d2}, the final distance between the particles is $d=3\lambda/4$.}
   \label{tFA}
\end{figure}

Very interesting physics can also be seen in the time evolution of the system, when we allow the particles to dynamically adjust according to the local forces by solving the coupled equations of fields and particle motion. We will introduce these equations in more detail later in the many particle section in Eq.~\eqref{Newtoneq}. In the simplest nontrivial case of two particles and $I_l=I_r=0$, we can use this to see how the system finds a stationary, self-consistent equilibrium. A typical example is shown in Fig.~\ref{tFA}a), where starting from an unstable zero force distance $d=\lambda/4$ the particles adjust to the above calculated equilibrium at $d=3 \lambda/4$. Interestingly in the final configuration they are not simply trapped near local intensity maxima but form a self-ordered optical resonator with intensity maximum at its center. 

A very similar behaviour is found in Fig.~\ref{tFA}b) where we consider strongly absorbing particles. Again we find almost the same equilibrium positions but with much less pronounced light confinement. Still it is a bit surprising that they are not pushed outwards by the confined light between them. This reminds of the self-ordered solutions found in continuous Vlasov model for an ultracold gas in such a field~\cite{griesser2013light}.  

\subsection{Three particles}
\begin{figure}
	\includegraphics[width=\columnwidth]{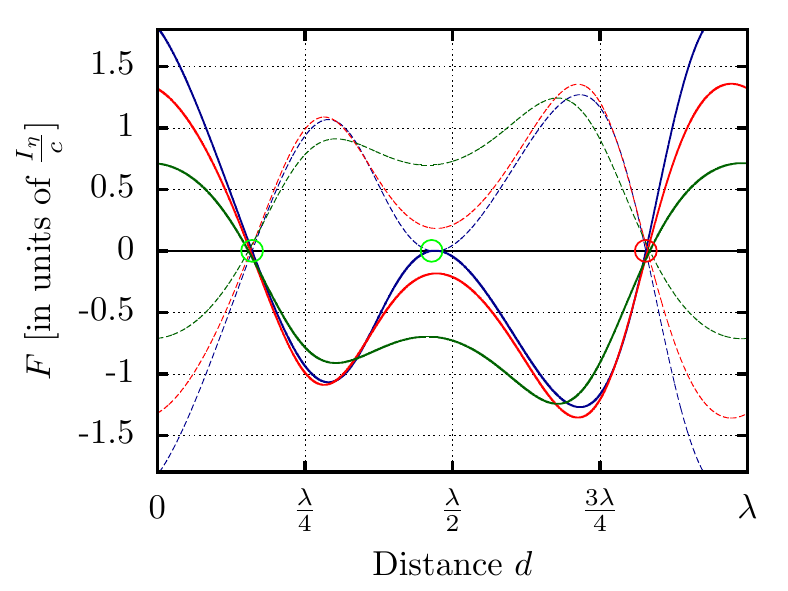}
	\caption{Force on three particles as function of the distances between the beam splitters $d=d_1=d_2$ for $I_l=I_r=0$, $\alpha=\beta$ and $\phi=0$. The solid line shows the force on the first beam splitter and the dashed line the force on the third one. The force on the second beam splitter is zero. The blue line corresponds to $\zeta=1/9$, the red line to $\zeta=(1+i)/9$ and the green line to $\zeta=1/9+i/2$. The red circles are stable points and the green ones not.}
\label{fig_F3}
\end{figure}

Let us now add a third particle but still no injected fields  $I_l=I_r=0$. While the general expressions for fields and forces can still be found explicitly, their form is already so complicated, that it is not instructive to print them here. Fortunately for a symmetric configuration  $d_1=d_2=d$ and neglecting absorption, $\zeta\in\mathbb{R}$, we still get the outgoing field in an instructive and useful form:
\begin{equation}
\begin{split}
	&I_{ol}=I_{or}=\\
	&\frac{I_\eta (1+\zeta^2)}{2}\left|\frac{ 1+2 \cos(k d)-2 \zeta \sin(k d))}{(i+3\zeta) \cos(k d)+(1-\zeta(i+2\zeta)) \sin(k d)}\right|^2 
	\end{split}
\end{equation}
 For small $\zeta$ this reduces to $I\simeq I_\eta /2 \left| 1+2\cos(k d))\right|^2$ which can lead up to 9 times stronger collective scattering than for a single particle. From symmetry considerations we again see that $F_2=0$ and the remaining forces sum up to zero (Fig.~\ref{fig_F3}) with
\begin{equation}
\begin{split}
  F_1=-F_3&\simeq\frac{I_\eta}{c}\left( (\cos(k d)+\cos(2 k d))\right.\\
  &\left.-\zeta  (2 \sin(2 k d)+\sin(3 k d)+\sin(4 k d))\right) + O[\zeta ]^2
\end{split}
\end{equation}

\begin{figure}
	\includegraphics[width=\columnwidth]{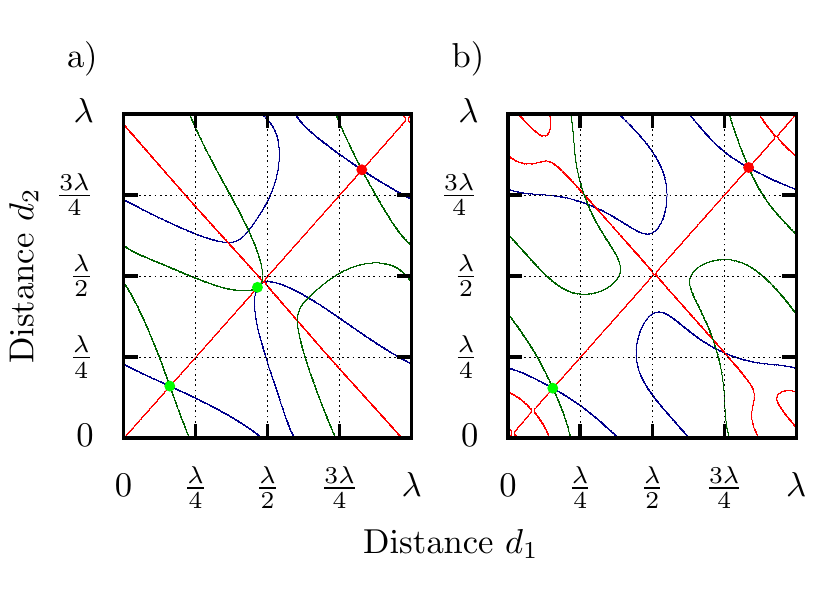}
	\caption{Contour lines of zero force $F_1$ (blue), $F_2$ (red) and $F_3$ (green) as function of the distance between the beam splitters one and two $d_1$ and the beam splitters two and three $d_2$ for $I_l=I_r=0$, $\alpha=\beta$ and $\phi=0$. The red points show stable points and the green ones unstable equilibria. Figure a): $\zeta=1/9$, Figure b): $\zeta=1/9+i/2$. It is interesting to see that one of the unstable zero points vanishes wehen we add an imaginary part to $\zeta$.}
	\label{force3c}
\end{figure}

In this symmetric case we only have stable configurations if $d_1=d_2\approx 4 \lambda/5$, which again are very insensitive to absorptive losses in the particles as shown in Fig.~\ref{force3c}.

A priori one might assume that also asymmetric solutions with different particle distances $d_1\neq d_2$ are possible. This can be checked by simply drawing zero force lines for all three particles as function of $d_1$ and $d_2$. Common intersections of all three lines then denote stationary configurations as shown in Fig.~\ref{force3c}. Their stability can be checked from the corresponding derivatives of the forces. We see that in the chosen examples stable equilibria only appear if $d_1 = d_2$, at least for symmetric scattering $\alpha=\beta$.

\section{Dynamics of larger particle ensembles}

In principle, adding more scatterers is straightforward in our model: we simply have to multiply by one more scattering matrix and find the solution of the corresponding set of linear equations for the field amplitudes. Explicit analytic results for the fields and forces can thus still be found, but these look surprisingly complex even for fairly small particle numbers. Finding the common zeros for all forces then is virtually impossible and we need to make rather drastic simplifications to arrive at useful analytical results. 

\subsection{Very weak coupling limit $\zeta =0$}
In their pioneering work, Chang and coworkers have found an elegant and surprisingly simple general result for two-level atoms in the weak excitation regime~\cite{chang2013self}. Here the induced dipoles interacting over infinite range via the fiber mode simply have to be arrange in a way such that the sum of the dipole couplings vanishes. This leads to a simple equidistant lattice configuration. In our model we can reproduce a similar limiting case by neglecting the coupling parameter $\zeta$ (i.e. setting $\zeta=0$), while still keeping nonzero scattering into the fiber $\eta \ne 0$. As both quantities in principle are proportional to the linear polarizability of the particle this is not possible in a strict sense, but a small $\eta$  can be compensated by using a much stronger pump laser, such that the ratio of the two scattering parameters $(\zeta,\eta)$ can be tuned to some extend. 

For $\zeta =0 $ the equations considerably simplify and in particular if we assume equidistant ordering $d_1=d_2=\dots=d_{N-1}=d$ the total intensities scattered into the fiber finally read
\begin{equation}
	I_{ol}=I_{or}=\frac{I_\eta}{2}\left(\frac{\sin(\frac{N k d}{2})}{\sin(\frac{k d}{2})}\right)^2
\end{equation}

\begin{figure}
\includegraphics[width=\columnwidth]{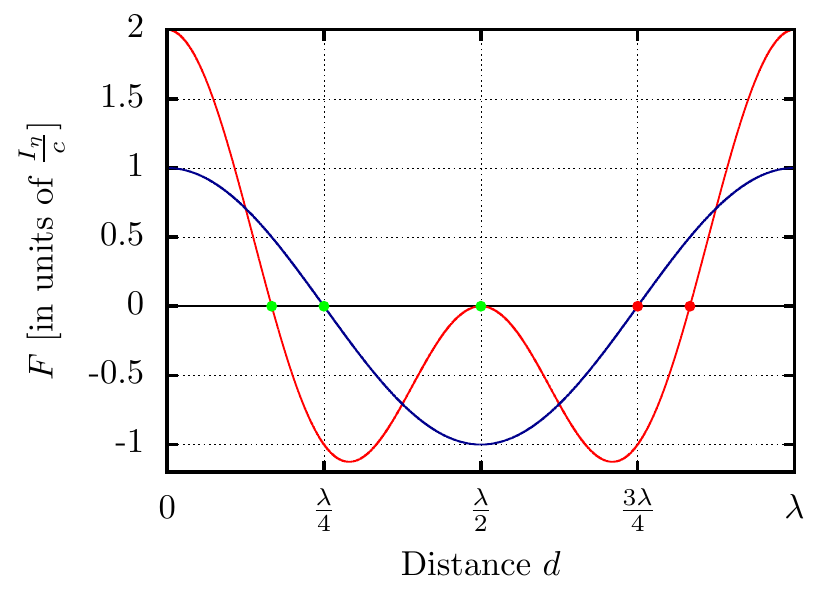}
\caption{Force on the first of two (blue) and three (red) particles for vanishing $\zeta=0$ and symmetric scattering $\alpha=\beta$. Red dots denote stable equilibrium points. It confirms that the stable points are at $d=(2N-1)/(2N)$.}
\label{force0}
\end{figure}

Also the force on the $m$-th particle of $N$ beam-splitters can be obtained in closed form to give
\begin{equation}
	F_m=-\frac{I_\eta  \cos\left(N k d/2\right)\sin\left((2 m-N-1)k d/2\right)}{c \sin\left(k d/2\right)} .
\end{equation}
The zeros of the force are then given by
\begin{equation}
	d=\frac{2n-1}{2 N}\lambda, \mbox{with} \  n\in\mathbb{N}
\end{equation}
predicting a regular, equally spaced distribution of the scatterers. This agrees well with the previous prediction but one still has to check for stability of this solutions. As symmetry again enforces a stationary center of mass,  we simply look at the leftmost of the N particles and try to identify its stable positions. In Fig.~\ref{force0} we plot the force on the leftmost of two and three particles as a function of distance. We see that the force vanishes at the points predicted above but only few zeros correspond to stable equilibria. Indeed we have to choose $n=N$ to guarantee stability.

\begin{figure}
\includegraphics[width=\columnwidth]{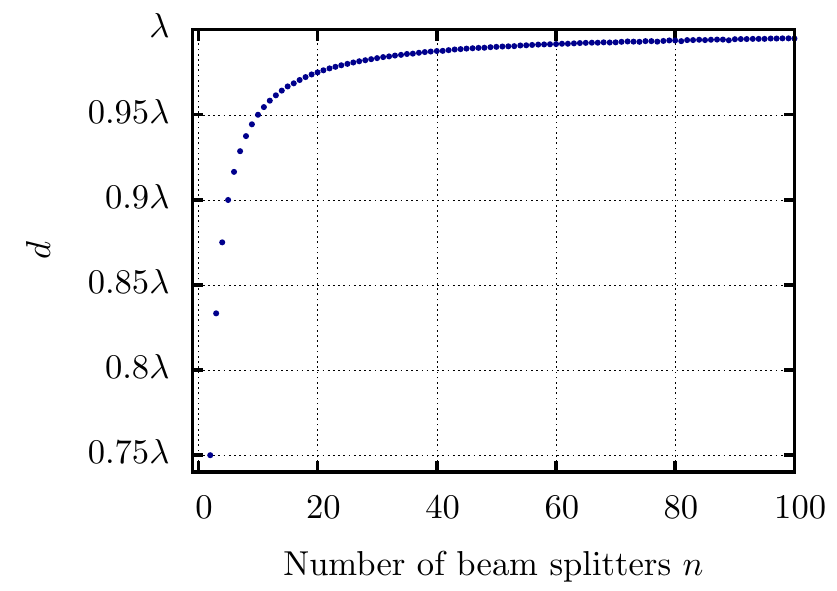}
\caption{Stable distance as function of number of beam splitters $N$, numerically solved with initial condition $d_0=\lambda$ for $\zeta=0$, $I_l=I_r=0$ and $\alpha=\beta$.}
\label{nd0}
\end{figure}

This behavior is confirmed by a numerical solution for the stable configuration closest to $d=\lambda$ as shown in Fig.~\ref{nd0} depicting the stable distance of the equally spaced distribution as a function of the number of scatterers confirming the rule found above $d=\frac{2N-1}{2N}\lambda$. This behaviour relates to the fact that in this limit every particle interacts equally strong with all the other particles via the unperturbed fiber mode. In the next section (see Fig.~\ref{dist_n_zeta}) we will show how this behaviour changes in the limit of small but finite $\zeta$ including absorption. 

\subsection{Numerical simulations for large ensembles}
While the general equations for the field evolution and forces can be written down fairly easily even for larger particle numbers, their solution gets very complex and hard to interpret even for small particle numbers. Nevertheless a numerical solution of the self-consistent dynamical Newtonian equations of motion for the particles with mass $m$ and friction coefficient $\mu$ can be easily performed until an equilibrium configuration is reached. 
We start from the classical equations of motions including some prescribed damping $m\ddot{x}_j=-\mu\dot{x}_j+F_j(x_1,\cdots,x_N)$, where the force is determined from the momentary field configuration. In an over-damped limit the velocity is determined by the force over friction ratio, so that we have:
\begin{equation}\label{Newtoneq}
 \dot{x}_j=\frac{F_j(x_1,\cdots,x_N)}{\mu}.
\end{equation}

\begin{figure}
\centering
 \includegraphics[width=\columnwidth]{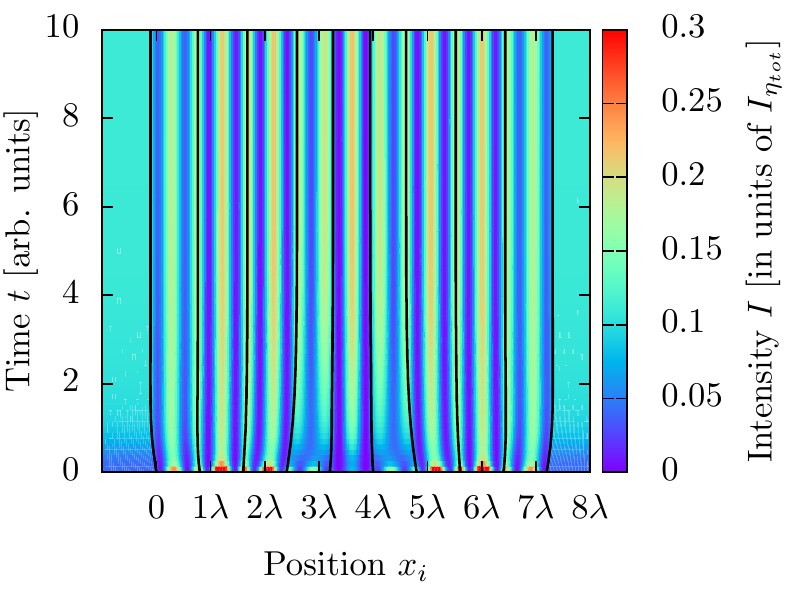} 
  \caption{Trajectories of ten beam splitters for $I_l=I_r=0$, $\alpha=\beta$ and $\phi=0$ with initial condition $d_1=d_2=\dots=d_{9}=0.8 \lambda$ and $\zeta=(1+i)/9$. Note that the outermost particles are not trapped at intensity nodes or antinodes as would be expected from a conventional lattice.}
  \label{tFAd}
\end{figure}

As the field adiabatically follows the particle distribution, the system can be expected to eventually evolve towards a self-consisted equilibrium position. In Fig.~\ref{tFAd} this is demonstrated at the example of ten particles initially prepared in an equidistant chain of distance $d=0.8\lambda$. Once the self-consistent dynamics are started, the particles redistribute to a new stable order with varying distance. Interestingly we again see that the outer particles are not drawn towards high field intensity regions but form a kind of resonator confining a great deal of the scattered light close to the particles. By maintaining a somewhat smaller seperation than the inner particle, the outer particles act as reflectors, which form a more conventional optical lattice trapping the inner particles close to field maxima. 

\begin{figure}
  \includegraphics[width=\columnwidth]{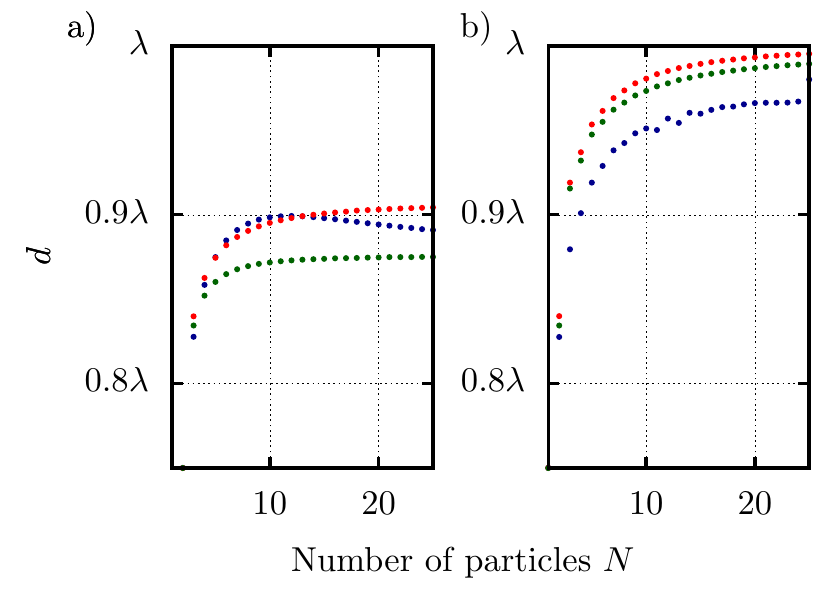} 
\caption{Stable distance $d_1$ (Figure a)) and $d_{N/2}$ for a even number, respectively $d_{(N+1)/2}$ for a odd number of particles (Figure b)) as function of number of beam splitters $N$, numerically solved with initial condition $d_0=\lambda$. The red line corresponds to $\zeta=1/9$, the blue line to $\zeta=i/2$ and the green line to $\zeta=1/9+i/2$.}
\label{dist_n_zeta}
\end{figure}

Changing the number of particles leads to a very similar behavior with slightly modified distances. In Fig.~\ref{dist_n_zeta} we plot the dependence of the distance of the outermost two particles and two particles in the middle of the stationary lattice as function of particle number. While the distance first grows with particle number, it reaches a stationary value with an effective lattice constant below one wavelength $\lambda$ for more than ten particles in the chain. Surprisingly even for the purely absorptive case with imaginary $\zeta$ a stable stationary order can be induced by the transverse pump light. 

Hence we see, that stable self-ordered lattices can form for large particle numbers, as long as some friction is present. This strongly resembles the results obtained via the Vlasov approach presented in ref.~\cite{griesser2013light}. Note that at the center we almost get a wavelength spaced optical lattice with particles trapped at field maxima, while the outer particles attain a smaller spacing forming an effective mirror for the light. The system hence acts like a self-forming optical resonator trapping the scattered light in its center. Such a configurations tends to minimize the combined total potential energy of all the particles~\cite{asboth2007comment}.      

\begin{figure}
\includegraphics[width=\columnwidth]{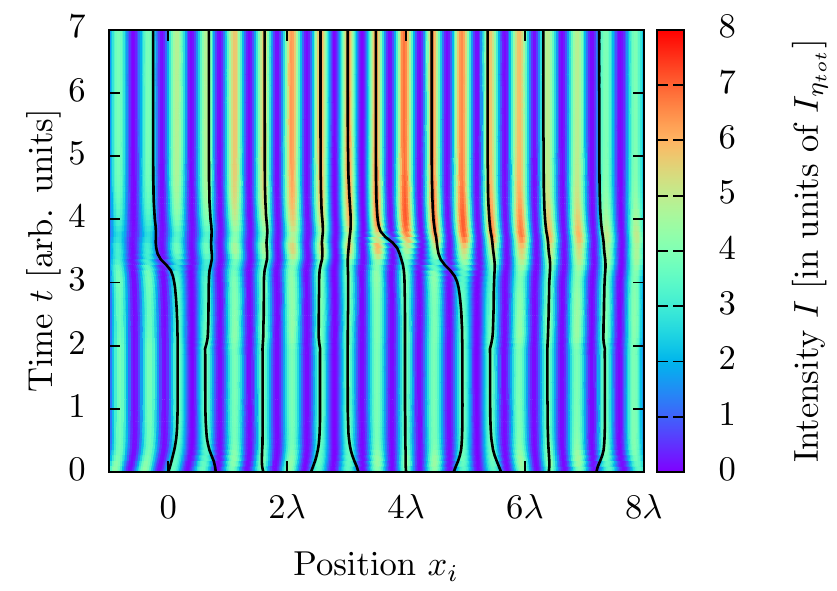}
\caption{Trajectories of ten beam splitters for $I_l=I_r=I$, $\zeta=1/9$, $\alpha=\beta$ and $\phi=0$ with initial condition $d_1=d_2=\dots=d_9=0.8 \lambda$. For the first three time-steps $I_{\eta_{tot}}=0$, then $I_{\eta_{tot}}=I$. We see that the particles order in a new stable configuration when we switch on the transverse pump at $t=3$.}
\label{tFAeta}
\end{figure}
To emphasize the difference between a typical 1D-lat\-tice and a lattice with transverse pump we present the effects of a sudden switch on of the transverse coupling $I_\eta$ in Fig.~\ref{tFAeta}. There we initially inject a common standing wave into the fiber to prepare an optical lattice. The transverse pump laser is then used to induce additional non-local couplings between the particles.  Clearly, when we switch on $I_\eta$, the particles start to interact differently and re-order into a new equilibrium. Interestingly, again the reordering tends to have more light confined within the structure via multiple scattering while the outer particles no longer rest at positions of maximum intensity.

\begin{figure}
\includegraphics[width=\columnwidth]{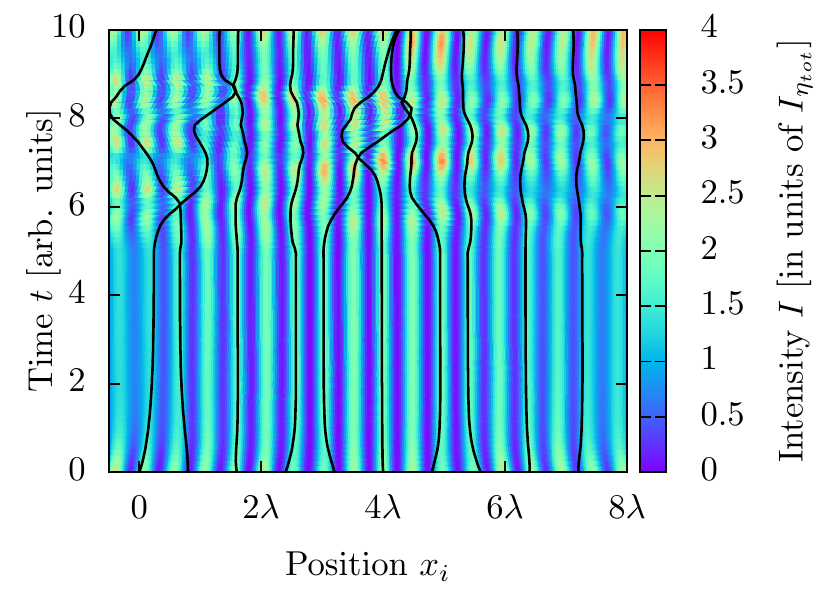}
\caption{Trajectories of ten beam splitters for $I_l=I_r=I$, $\zeta=1/9+i/18$, $\alpha=\beta$ and $\phi=0$ with initial condition $d_1=d_2=\dots=d_9=0.8 \lambda$. For the five time-steps $I_{\eta_{tot}}=0$, then $I_{\eta_{tot}}=I$. As the particle trajectories begin to cross we see that the configuration gets unstable one the transvere pump is switched on.}
\label{tFAeta1}
\end{figure}

In some cases, however, in particular for a complex $\zeta$ involving absorption, the effective interaction can be too strong to generate a new order and the lattice disintegrates after switching the transverse pump on even in the over-damped regime.  Such a behavior as shown in Fig.~\ref{tFAeta1}  was also found for very large conventional optical lattices and clearly demonstrates that we do not have conservative dynamics here~\cite{asboth2008optomechanical}. 

\section{Selfordering with asymmetric directional scattering amplitues $\alpha\neq\beta$}
Fiber and pump laser in our model constitute a translationally invariant system with two equivalent propagation directions and scattering between forward and backward direction will be symmetric at first glance. However, using transverse pump with a polarization, which is not aligned perpendicular to the fiber,  this symmetry is broken.  As recently outlined in ref.~\cite{mitsch2014directional}  in some cases this introduces directional scattering, an effect which can be magnified close to the fiber surface. To effectively model this behavior for general directions of pump polarization, we parametrize the part of $\eta$ reflected to the left as $\alpha=\sin(\theta)$ and the one reflected to the right $\beta=\cos(\theta)$, so that $\alpha^2+\beta^2=1$. In all previous examples we had $\alpha=\beta=1/\sqrt{2}$ or $\theta=\pi/4$. In the following we will shortly discuss the consequences of such an asymmetry on collective scattering, forces and particle ordering in some special examples. 

\begin{figure}
\includegraphics[width=\columnwidth]{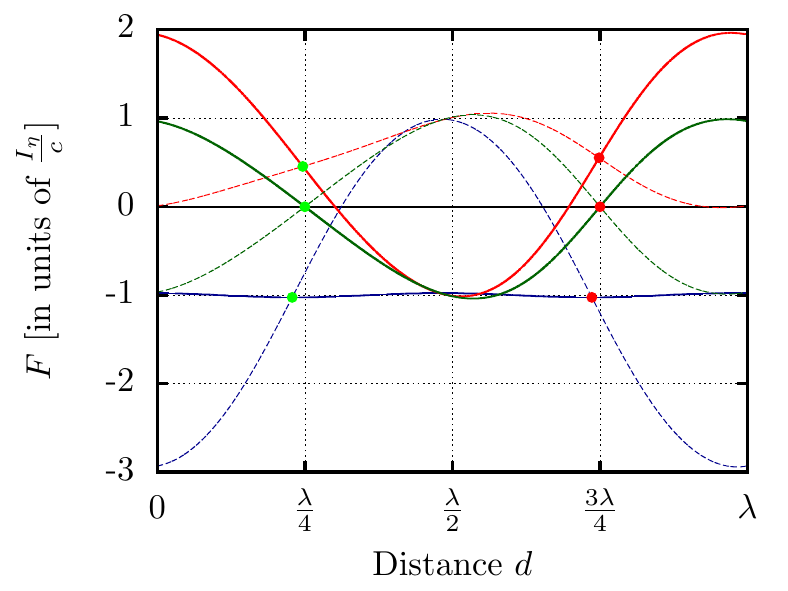}
\caption{Force on two beam splitters as function of the distance $d$ for $I_l=I_r=0$, $\zeta=1/9$ and $\phi=0$ with scattering asymmetries. The blue line corresponds to $\theta=0$, the red line to $\theta=\pi/3$ and the green line to $\theta=\pi/4$. The solid line shows the force on the first beam splitter and the dashed line the force on the second beam splitter.  Crossings of the lines with equal force on the two particles here also occur a nonzero force values. The red points are stable distances, while the green ones are not.}
\label{F2ab}
\end{figure}

For a single beam splitter the outgoing field intensities then change to:

\begin{equation}
\begin{split}
	&I_{ol}=\left|\frac{\left(\sqrt{I_r}+\sqrt{I_l} e^{i (\phi+2 k  x)} i \zeta\right)+e^{ i k  x}\sqrt{I_\eta} (1-i\zeta) \sin(\theta)}{1-i\zeta}\right|^2\\
	&I_{or}=\left|\frac{\left(\sqrt{I_l} e^{i (\phi+2 k x)}+ \sqrt{I_r} i\zeta\right)+e^{i k  x}\sqrt{I_\eta} (1-i\zeta) \cos(\theta)}{1-i\zeta}\right|^2
\end{split}
\end{equation}

From Eq.~\eqref{eq_force_singleBS_general} we see that a single particle is always pushed in one direction of $ I_r=I_l=0$. But a longitudinal pump or the presence of other beam splitters allows one to find stable configurations even if $\alpha\neq \beta$. For different asymmetry $\theta$ the number of zero points per wavelength can change.

\begin{figure}
  \includegraphics[width=\columnwidth]{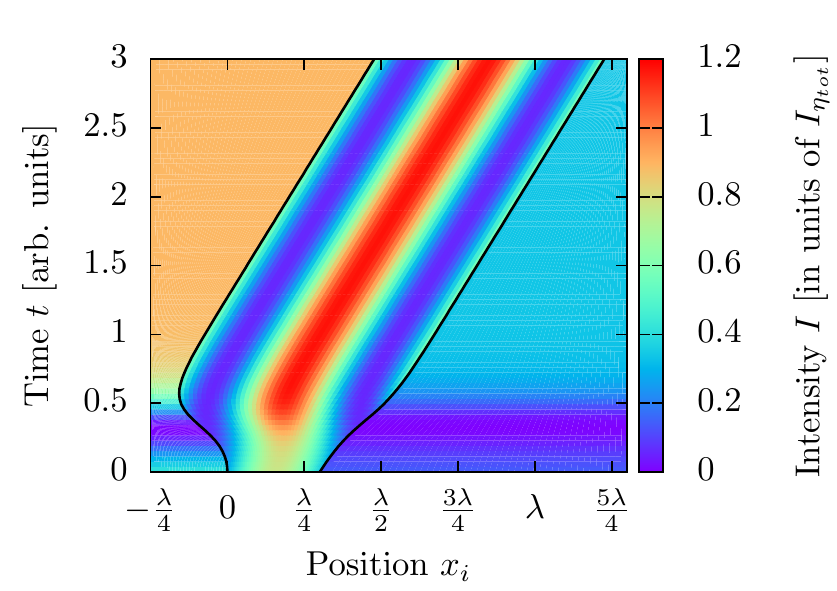}
  \caption{Trajectories of two beam splitters for $I_l=I_r=0$, $\zeta=1/9$, $\theta=\pi/3$ and $\phi=0$ with initial condition $d=0.3 \lambda$. We see that the particles are pushed to the right because of the scattering asymmetry, but they keep a stable final distance.}
  \label{tFAaf}
\end{figure}

Obviously such a scattering asymmetry also changes the interaction properties in the mulitparticle case. This has the important consequence that the forces on two beam splitters are not always of equal magnitude and typically a final force on the center of mass remains. Nevertheless it can lead to a stable distance of two beam splitters, where the force on the two particles is equal. This generates thus a propulsion of a pair of particles keeping an equal distance similar to experiments with silicon beads~\cite{frawley2014selective}. Examples of this behavior are shown in Figure~\ref{F2ab}, where we show the forces on a pair of scatterers as function of distance and scattering asymmetry $\theta$. For asymmetric scattering a finite net force remains at the stable intersection points (red dots) and even for perfect unidirectional scattering the two particles can be locked to a stable distance.

\begin{figure}
  \includegraphics[width=\columnwidth]{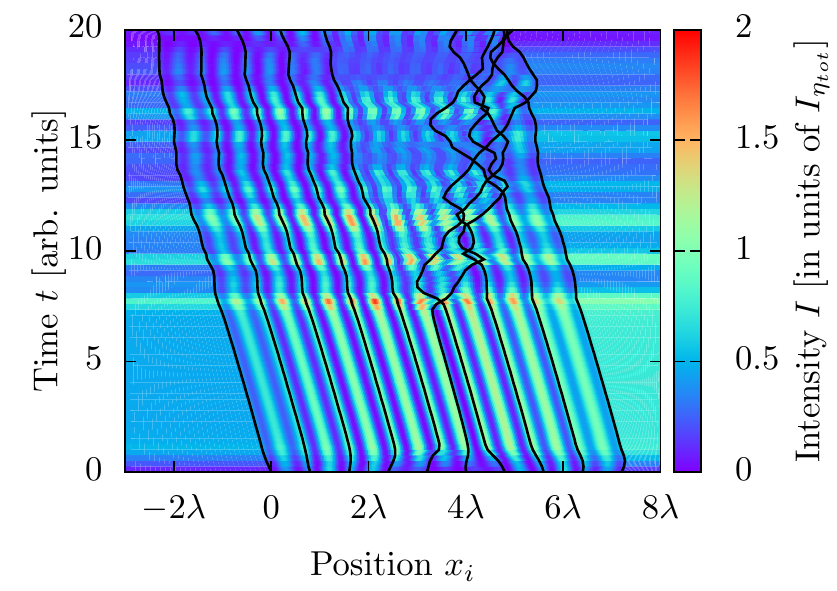}
   \caption{Trajectories of ten beam splitters for $I_l=I_r=0$, $\zeta=\frac{1}{9}$, $\theta=\pi/5$ and $\phi=0$ with initial condition $d_1=d_2=\dots=d_9=0.8 \lambda$.}
   \label{fig_10particles_unstable}
\end{figure}

Such a behavior is also shown in the dynamical solution of the asymmetric two-particle problem in Figure~\ref{tFAaf}. While the asymmetric scattering drives the particles in a preferred direction, we find stable parallel trajectories for the two particles. Interestingly, as in the symmetric case, the particles assume a distance, where they confine much of the scattered light between them. This is even more pronounced for larger particle numbers as shown in Fig.~\ref{fig_10particles_unstable}, where we exhibit such a more complex behavior for ten beam splitters. There we see that the distances starts to oscillate around a stable point and the system gets unstable after a while. Looking again at Fig.~\ref{F2ab}  we can trace this instability to the fact that the derivative of the force at the zero point is smaller if $\alpha\neq\beta$, corresponding to weaker particle distance locking.

Here we see that the most critical position is the outermost beamsplitter on the right, where weak perturbations are enough to decouple this particle from the rest. In this case also no resonator for the light is formed.

\section{Conclusions and outlook}

Using a simple classical scattering model we were able to study the collective dynamics and self-ordering of illuminated point scatterers coupled to the traveling wave fields along a tapered optical nano-fiber. This simple approach can reproduce the dipole-dipole induced self-ordering predicted for the very weak dipole-dipole coupling limit~\cite{chang2013self} as well as stable ordering and strong light confinement found for a cold gas mean-field approach~\cite{griesser2013light}. Our model naturally allows to include back-scattering and absorption and gives an intuitive picture for the underlying microscopic dynamics. It turns out that interference of the light scattered by one particle with the light coming from other particles constitutes the dominant contribution to the force. In an ideal fiber this interference happens at all distances and thus mediates interactions throughout the whole ensemble. For  large ensembles this can create strong instabilities such that the whole lattices structure disintegrates despite strong damping of particle motion.     

While the model is easy to formulate and in principle allows for an analytic treatment, the explicit expressions get quite complex and uninstructive even for small particle numbers. Numerical treatments are, however, straightforward and possible for very large particle numbers as the computational effort only slowly grows with particle number. Here we introduced some external friction force which allows one to identify stationary solutions, which in an experiment could be provided by Doppler cooling or similar mechanisms.

In contrast to a prescribed optical lattice the dynamics are not conservative and the particles will in general not occupy positions at field maxima corresponding to optical potential minima, but will order according to zero force configurations. This leads to unexpected and intriguing solutions, where the outer particles form a self-organized resonantor confining a great deal of the scattered light within the structure. This is similar to previous ideas of particle-based resonators but happens without the need to artifically fix the particles~\cite{chang2012cavity}. In contrast to the longitudinal case such a configuration is not intrinsically unstable~\cite{asboth2007comment} as the destructive interference of scattered and propagating fields can stabilize the outermost particles.  

Let us remark that the presentation and argumentation here was heavily based on the example of a nano-fiber, but generalization to other 1D cases as hollow core fibers or other field confining nano-structures are straightforward. In fact, if the particles are confined in a 1D geometry at sufficient density no auxiliary optical structure is required in principle, as they will also guide the light. Hence a sufficiently dense elongated atomic ensemble can be expected to spontaneously crystallize under coherent illumination forming a self-contained optical lattice. Interestingly such a lattice has intrinsically built-in long-range interactions and phononic degrees of motion, without the need of any optical resonators or auxiliary particles to mediate interactions. At least qualitatively such a behaviour can be expected for 2D particle confinement, where preliminary simulations hint for a hexagonal order with spatially slowly varying lattice constant. It is not clear how pumping of this kind could be implemented in 3D but maybe optical gain could pave a way to a 3D selfordered atom-light crystal.\\

{\bf Acknowledgments:}{We thank Arno Rauschenbeutel and Stefan Ostermann for helpful discussions. This work has been supported by the  by the Austrian Science Fund (FWF)through SFB Foqus project F4006 N16}

\bibliographystyle{osajnl}
\bibliography{scatt1d}

\begin{thebibliography}{10}
\newcommand{\enquote}[1]{``#1''}
\expandafter\ifx\csname url\endcsname\relax
  \def\url#1{\texttt{#1}}\fi
\expandafter\ifx\csname urlprefix\endcsname\relax\def\urlprefix{URL }\fi
\providecommand{\eprint}[2][]{\url{#2}}

\bibitem{deutsch1995photonic}
I.~Deutsch, R.~Spreeuw, S.~Rolston, and W.~Phillips, \enquote{Photonic band
  gaps in optical lattices,} Physical Review A \textbf{52}(2), 1394 (1995).

\bibitem{asboth2008optomechanical}
J.~Asb{\'o}th, H.~Ritsch, and P.~Domokos, \enquote{Optomechanical coupling in a
  one-dimensional optical lattice,} Physical Review A \textbf{77}(6), 063,424
  (2008).

\bibitem{vetsch2010optical}
E.~Vetsch, D.~Reitz, G.~Sagu{\'e}, R.~Schmidt, S.~Dawkins, and
  A.~Rauschenbeutel, \enquote{Optical interface created by laser-cooled atoms
  trapped in the evanescent field surrounding an optical nanofiber,} Physical
  review letters \textbf{104}(20), 203,603 (2010).

\bibitem{domokos2002quantum}
P.~Domokos, P.~Horak, and H.~Ritsch, \enquote{Quantum description of
  light-pulse scattering on a single atom in waveguides,} Physical Review A
  \textbf{65}(3), 033,832 (2002).

\bibitem{goban2012demonstration}
A.~Goban, K.~Choi, D.~Alton, D.~Ding, C.~Lacro{\^u}te, M.~Pototschnig,
  T.~Thiele, N.~Stern, and H.~Kimble, \enquote{Demonstration of a
  state-insensitive, compensated nanofiber trap,} Physical Review Letters
  \textbf{109}(3), 33,603 (2012).

\bibitem{lee2013integrated}
J.~Lee, D.~Park, S.~Mittal, M.~Dagenais, and S.~Rolston, \enquote{Integrated
  Optical Dipole Trap for Cold Neutral Atoms with an Optical Waveguide
  Coupler,} arXiv preprint arXiv:1303.2922  (2013).

\bibitem{zoubi2010hybrid}
H.~Zoubi and H.~Ritsch, \enquote{Hybrid quantum system of a nanofiber mode
  coupled to two chains of optically trapped atoms,} New Journal of Physics
  \textbf{12}(10), 103,014 (2010).

\bibitem{frawley2014selective}
M.~C. Frawley, I.~Gusachenko, V.~G. Truong, M.~Sergides, and S.~N. Chormaic,
  \enquote{Selective particle trapping and optical binding in the evanescent
  field of an optical nanofiber,} arXiv preprint arXiv:1403.7599  (2014).

\bibitem{asboth2005self}
J.~Asb{\'o}th, P.~Domokos, H.~Ritsch, and A.~Vukics, \enquote{Self-organization
  of atoms in a cavity field: Threshold, bistability, and scaling laws,}
  Physical Review A \textbf{72}(5), 053,417 (2005).

\bibitem{chan2003observation}
H.~W. Chan, A.~T. Black, and V.~Vuleti{\'c}, \enquote{Observation of
  collective-emission-induced cooling of atoms in an optical cavity,} Physical
  review letters \textbf{90}(6), 063,003 (2003).

\bibitem{chang2012cavity}
D.~Chang, L.~Jiang, A.~Gorshkov, and H.~Kimble, \enquote{Cavity QED with atomic
  mirrors,} New Journal of Physics \textbf{14}(6), 063,003 (2012).

\bibitem{griesser2013light}
T.~Grie{\ss}er and H.~Ritsch, \enquote{Light-induced crystallization of cold
  atoms in a 1D optical trap,} Physical review letters \textbf{111}(5), 055,702
  (2013).

\bibitem{mitsch2014discerning}
R.~Mitsch, C.~Sayrin, B.~Albrecht, P.~Schneeweiss, and A.~Rauschenbeutel,
  \enquote{Discerning and selectively manipulating laser-trapped atoms using
  non-paraxial light,} arXiv preprint arXiv:1403.5129  (2014).

\bibitem{sonnleitner2012optomechanical}
M.~Sonnleitner, M.~Ritsch-Marte, and H.~Ritsch, \enquote{Optomechanical
  deformation and strain in elastic dielectrics,} New Journal of Physics
  \textbf{14}(10), 103,011 (2012).

\bibitem{ostermann2014scattering}
S.~Ostermann, M.~Sonnleitner, and H.~Ritsch, \enquote{Scattering approach to
  two-colour light forces and self-ordering of polarizable particles,} New
  Journal of Physics \textbf{16}(4), 043,017 (2014).

\bibitem{xuereb2009scattering}
A.~Xuereb, P.~Domokos, J.~Asb{\'o}th, P.~Horak, and T.~Freegarde,
  \enquote{Scattering theory of cooling and heating in optomechanical systems,}
  Physical Review A \textbf{79}(5), 053,810 (2009).

\bibitem{mitsch2014directional}
R.~Mitsch, C.~Sayrin, B.~Albrecht, P.~Schneeweiss, and A.~Rauschenbeutel,
  \enquote{Directional nanophotonic atom--waveguide interface based on
  spin-orbit coupling of light,} arXiv preprint arXiv:1406.0896  (2014).

\bibitem{chang2013self}
D.~E. Chang, J.~I. Cirac, and H.~J. Kimble, \enquote{Self-Organization of Atoms
  along a Nanophotonic Waveguide,} Phys. Rev. Lett. \textbf{110}, 113,606
  (2013).

\bibitem{asboth2007comment}
J.~Asboth and P.~Domokos, \enquote{Comment on “Coupled dynamics of atoms and
  radiation-pressure-driven interferometers” and “Superstrong coupling
  regime of cavity quantum electrodynamics”,} Physical Review A
  \textbf{76}(5), 057,801 (2007).

\end{thebibliography}
\end{document}